\documentclass[12pt,aaspp4]{aastex}
\slugcomment{{\em}}

\begin{document}

\title{Statistical Properties of Collisionless Equal- and Unequal-Mass Merger Remnants of Disk Galaxies}

\author{Thorsten Naab}
\affil{Institute of Astronomy, Cambridge University, Madingley Road,
Cambridge, CB3 0HA}
\affil{Max-Planck-Institut f\"ur Astronomie, K\"onigstuhl 17, 69117
Heidelberg}
\author{Andreas Burkert}
\affil{Max-Planck-Institut f\"ur Astronomie, K\"onigstuhl 17, 69117
Heidelberg}

\begin{abstract}
We perform a large parameter survey of collisionless N-body
simulations of binary mergers of disk-galaxies with mass ratios of 1:1, 2:1,
3:1, and 4:1 using the special purpose hardware GRAPE. A set of 112
merger simulations is used to investigate the fundamental statistical
properties of merger remnants as a function of the initial orientation and
the mass ratio of the progenitor disks. The photometric and
kinematical properties of the simulated merger remnants are
analyzed. The methods used to determine the characteristic properties
are equivalent to the methods used for observations of giant
elliptical galaxies. We take projection effects into account and
analyze the remnant properties in a statistical way for comparison
with observations. The basic properties of the remnants correlate with
the mass ratio of the progenitor disks. We find that about 80\% of
the equal-mass merger simulations lead to slowly rotating  merger
remnants having $(v/\sigma)^* < 0.4$. Observers would interpret
those objects as being supported by anisotropic velocity
dispersions. All 1:1 remnants show significant minor-axis
rotation. One half of all projected 1:1 remnants shows boxy ($a_4 < 0
$) and the other half shows disky  
($a_4 > 0 $) isophotes. A distinct subclass of 4 out of 12 initial
orientations leads to purely boxy remnants independent of
orientation. 1:1 mergers with other initial orientations show disky or
boxy isophotes  depending on the viewing angle. Remnants with mass
ratios of 3:1 and 4:1 have more homogeneous properties. They all
rotate fast (maximum value of $(v/\sigma)= 1.2$) and show a small
amount of minor-axis rotation, consistent 
with models of isotropic or slightly anisotropic oblate rotators. If 
observed in projection they would be interpreted as being supported by
rotation. About 90\% of the projected 3:1 and 4:1 remnants show disky 
isophotes. 2:1 remnants show intermediate properties. Projection
effects lead to a large spread in the data in good agreement with 
observations. They do not change the fundamental kinematical
differences between equal-mass and unequal-mass merger remnants.   
The correlation between isophotal twist and apparent ellipticity of
every single merger remnant is in good agreement with
observations. The amount of twisting strongly depends on the  
orientation of the remnant but is only weakly dependent on the mass
ratio of the merger.  The results of this study weaken the disk merger
scenario as the possible formation mechanism of massive boxy giant
ellipticals as only equal-mass mergers with special initial orientations
can produce purely boxy anisotropic merger remnants. Some orientations
of 1:1 mergers can even lead to disky and anisotropic remnants which
are either not observed or would be classified as S0s based on their
morphology. In general, the properties of equal-mass (and 2:1) merger
remnants are consistent with the observed population of giant
ellipticals in the  intermediate mass regime between low mass fast
rotating disky and bright massive boxy giant ellipticals. 3:1 and 4:1
merger remnants, however, are in very good agreement with the class of
low luminosity, fast rotating giant elliptical galaxies. Binary
mergers of disk galaxies are therefore still very good candidates for
being the main formation mechanism for intermediate and low mass giant
ellipticals. The homogeneous class of massive boxy ellipticals most
likely formed by a different process.    

\end{abstract}

\keywords{galaxies: interaction-- galaxies: structure -- galaxies:
evolution -- galaxies: ellipticals -- methods: numerical }

\section{Introduction}
Detailed observations of individual giant elliptical galaxies have shown
that they can be subdivided into two groups with respect to their
structural properties \citep{bdm1988, b1988, kb1996} [BDM, and references
therein]. Faint giant ellipticals are isotropic rotators with small minor
axis rotation and  disky deviations of their isophotal contours from
perfect ellipses. They might contain faint disks which contribute up to 30\%
to the total light in the galaxy. Therefore their disk-to-bulge ratios
overlap with those of S0-galaxies \citep{rw1990,sb1995}. Disky
ellipticals have power-law inner density profiles \citep{l1995,f1997}
and show little or no radio and X-ray emission \citep{bsdm1989}. Boxy
ellipticals, on the other hand, are in general more massive than disky
ellipticals. They have box-shaped isophotes, and show flat cores. The
kinematics of boxy ellipticals is generally more complex than that of
disky ellipticals. They rotate slowly, are supposed to be supported by
velocity anisotropy and have a large amount of minor axis rotation. A
number of rotationally supported ellipticals also show boxy
isophotes. These systems are either purely boxy (this is supposed  
to originate from tidal interactions with nearby massive companions)
or show disky isophotes in the inner part and boxy isophpotes in the
outer part \citep{nb1989}. Occasionally, boxy ellipticals have
kinematically distinct cores \citep{fi1988,js1988,b1988a}. These cores 
inhabit flattened rapidly rotating disk- or torus-like components
dominating the light in the central few hundred parsecs
\citep{b1990,rw1992,m1998}, but they contribute only a few percent to
the total light of the galaxy. The fact that these cores are
metal-enhanced shows that gas (at least in the inner regions) must
have played an important role during their formation
\citep{bs1992,dsp1993,b1996,d1996}. Boxy ellipticals show stronger 
radio emission than the average and have high X-ray luminosities
consistent with emission from hot gaseous halos \citep{beu1999}. The
distinct physical properties of disky and boxy elliptical galaxies
point to the fact that both types of ellipticals could have different
formation histories. It has been argued by \citet{kb1996}
and \citet{f1997} that the observed stellar disks and the high density
power law centers in disky ellipticals are signatures of dissipation
during their formation. In this sense they seem to continue the 
Hubble sequence from S0s to higher bulge-to-disk ratios
\citep{kb1996}. Boxy ellipticals with an even higher bulge-to-disk
ratio show a stronger kinematical decoupling at their centers and no
signature of a disk at all. Therefore the early type Hubble sequence
from S0s to massive ellipticals might represent galaxies with a disk
component embedded in more prominent spheroidal body.\\  

The traditional view on the formation and evolution of giant
elliptical galaxies is that they are very old stellar systems and
formed very early at a redshift of more than two 
\citep{ssb1973}. After an intensive initial star-formation phase they
experienced very little mass evolution \citep{bc1993}. It has been
argued by many authors that the stellar evolution of ellipticals is
compatible with pure passive evolution models
\citep{bower1992,aragon1993,bzb1996,ellis1997,zb1997} or models with
exponentially decaying star formation \citep{z1999}.
Alternatively, hierarchical theories of galaxy formation predict that
massive galaxies were assembled relatively late in many generations of
mergers of disk-type galaxies or smaller subunits and mass accretion.
It has been argued by \citet{k1996} and \citet{kc1998} that this merger
scenario is consistent with observations of galaxies at different
redshifts. The idea that elliptical galaxies can form from mergers of
disk-galaxies has been originally proposed by
\citet{tt1972}. Thereafter the ``merger hypothesis``  has been
investigated in great details by many authors. For a recent review of
current models of spheroid formation see
\citet{BN2003a,BN2003b}. \citet{nw1983}, \citet{bar1988} and 
\citet{her1992} performed the first fully self-consistent merger
models of two equal-mass stellar disks embedded in dark matter
halos. The remnants were slowly rotating, pressure supported and
anisotropic systems and generally followed an $r^{1/4}$ surface density
profile in the outer parts. However, due to phase space limitations
\citep{c1986} it was necessary to start with progenitors with  massive
central bulge components  \citep{her1993b} to fit the observed de
Vaucouleurs profile also in the inner parts. These simulations showed
that global properties of equal-mass merger remnants resemble those of
ordinary slowly rotating massive elliptical galaxies. More detailed
investigations of isophotal 
shapes of the merger remnants have shown that the same remnant can
appear either disky or boxy when viewed from different directions
\citep{her1993b} with a trend pointing to boxy isophotes
\citep{hhs1994,sm1995}. \citet{b1998} and \citet{BB2000} investigated a 
sample of disk-disk mergers with a mass ratio of 3:1 and found that the
remnants are flattened and fast rotating in contrast to equal-mass
mergers. \citet{nbh1999} investigated the photometric and kinematical
properties of a prototypical 3:1 merger remnant in details and
compared the results to observational data of disky elliptical
galaxies. They found an excellent agreement and proposed that fast
rotating disky elliptical galaxies can originate from pure  
collisionless 3:1 mergers while slowly rotating, pressure supported
ellipticals form from equal-mass mergers of disk galaxies. In this
paper we extend the analysis of \citet{nbh1999}. A large number of
112 merger remnants resulting from a statistically unbiased sample of
simulations of mergers between disk-galaxies with mass ratios of 1:1,
2:1, 3:1, and 4:1 is investigated. This large sample allows a much
more thorough investigation of the statistical properties of
merger remnants as all previous studies. After a short description of
the simulation methods in section \ref{mmod} we investigate the
photometric and kinematic properties of the simulated remnants in
section \ref{pkp}. We discuss the drawback of our results on the
theory of the formation of elliptical galaxies in section \ref{con}.

\section{The merger models}\label{mmod}
The spiral galaxies were constructed in dynamical equilibrium
using the method described by \citet{her1993a}. We used the following
system of units: gravitational constant G=1, exponential scale length
of the larger disk $h=1$ and mass of the larger disk $M_d=1$. 
Each galaxy consists of an exponential disk, a spherical, non-rotating
bulge with mass $M_b = 1/3$, a Hernquist density profile
\citep{her1990} with a scale length $r_b=0.2h$  and a spherical
pseudo-isothermal halo with a mass $M_d=5.8$, cut-off radius $r_c=10h$
and core radius $\gamma=1h$. 

We followed a sequence of mass ratios of the progenitor disks from
$\eta =1$ to $\eta =4$ where $\eta $ is the mass of the more massive
galaxy divided by the mass of the merger partner. The equal-mass
mergers were calculated adopting in total 400000 particles with each
galaxy consisting of 20000 bulge particles, 60000 disk particles, and
120000 halo particles.  We decided to use twice as many halo particles   
than disk particles to reduce heating and instability effects in the
disk components \citep{nbh1999}. For the mergers with $\eta =2,3,4$
the parameters of the more massive galaxy were as described above. The
low-mass companion contained a fraction of $1 / \eta $ the mass and
the number of particles in each component with a disk scale length of
$h=\sqrt{1/\eta }$, as expected from the Tully-Fisher relation
\citep{PT1992}. 

The N-body simulations for the equal-mass mergers were performed by
direct summation of the forces using the special purpose hardware
GRAPE6 \citep{mfn2003}. With this highly efficient hardware one force
calculation for 400000 particles takes approx. $11$ seconds. The mergers
with mass ratios $\eta =2,3,4$ were followed using the newly developed
treecode VINE (Wetzstein et al., 2003) in combination with the GRAPE5
\citep{kfm2000} hardware for which the code was optimized. VINE uses a
binary tree in combination with the refined multipole acceptance
criterion proposed by \citet{ws1995}. This criterion takes the mass
distribution of every node into account. It enables the user to
control the absolute force error which is introduced by the tree 
construction. We chose a value of 0.001 which guarantees that the
error resulting from the tree is smaller than the intrinsic force error of
the GRAPE5 hardware which is of the order of $0.1\%$. To compensate
for the limited mass resolution of GRAPE5 we limit the maximum mass of
a tree node to $10^3$ times the minimum particle mass. To further
increase the speed of the calculation we only use GRAPE5 if forces
for more than 50 particles are needed at the same time. Otherwise we
compute the forces on the host computer as for small particle numbers
the communication time between the board and the host computer exceeds
the time for the force calculation on the host. With the parameters given
above one force calculation with VINE and GRAPE5 for 400000 particles
takes approx. $12$ seconds.  

We used a gravitational Plummer-softening of
$\epsilon = 0.05$ and a fixed leap-frog integration time step of
$\Delta t = 0.04$. For the equal-mass mergers simulated with direct
summation on GRAPE6 the total energy is conserved, VINE in
combination with GRAPE5 conserves the total energy up to $0.5\%$. 

The galaxies approached each other on nearly
parabolic orbits with an initial separation of $r_{sep} = 30$ length
units and a pericenter distance of $r_p = 2$ length units (same
parameters as e.g. \citet{her1992}). A study of orbits of merging dark
matter halos in cosmological large scale simulations by
\cite{koch2003} has shown that most of the merging halos are indeed on
parabolic orbits.   
The inclinations of the two disks relative to the orbit plane were $i_1$
and $i_2$ with arguments of pericenter $\omega_1$ and $ \omega_2
$. In selecting unbiased initial parameters for the disk inclinations
we followed the procedure described by \citet{b1998}. For the spin
vector of each disk we defined four different orientations pointing to
every vertex of a regular tetrahedron. The initial orientations we
used translate to the following set of angles: For the first galaxy
$i_1 = (0,-109,-109,-109), \omega_1 = (0,-60,180,60)$. The second
galaxy has $i_2 = (180,71,71,71), \omega_2 = (0,-30,30,90)$. These
parameters result in 16 initial configurations for equal-mass mergers
and 16 more for every mass ratio $\eta =2,3,4$ where the initial
orientations are interchanged. Following the simple hypothesis that
the orientations of the merging disks are independent of each other
and independent of their mutual orbital plane, every merger
geometry has an equal probability to be realized
\citep{b1998}. The orbital parameters are listed in Table
\ref{tbl-1}. In total we simulated 112 mergers. The total computing
time for the entire sample was about 1600 hours wall clock time.

For all simulations the merger remnants were allowed to settle into
equilibrium approximately  8 to 10 dynamical times after the
merger was complete. Then their equilibrium state was analyzed.

\section{Photometric and kinematical properties of the remnants}
\label{pkp}
To compare our simulated merger remnants with observations we analyzed
the remnants with respect to observed global photometric and
kinematical properties of giant elliptical galaxies, e.g. surface 
density profiles, isophotal deviation from perfect ellipses, velocity
dispersion, and major- and minor-axis rotation. Defining characteristic
values for each projected remnant we followed as closely as possible the
analysis described by BDM.

\subsection{Isophotal shape}
An artificial image of the remnant was created by binning the central 
10 length units into $128 \times 128$ pixels. This picture was
smoothed with a Gaussian filter of standard deviation 1.5 pixels. The
isophotes and their deviations from perfect ellipses were then
determined using a data reduction package kindly provided by Ralf
Bender. 

To be confident that a once determined isophotal shape is
characteristic for the remnant and does not change with time we
investigated the time evolution of the ellipticity and the $a_4$
profile starting $\approx 20$ time units after the merger of 
the bulge components was complete and followed the evolution for 
the next $50$ time units.  In intervals of four time units the
luminous part of the remnant was transformed to the principal axes of 
its moment of inertia tensor. The tensor was evaluated using $40\%$ of
the most tightly bound particles. The isophotal properties were then
analyzed as seen along the minor axis

\begin{figure}
\plotone{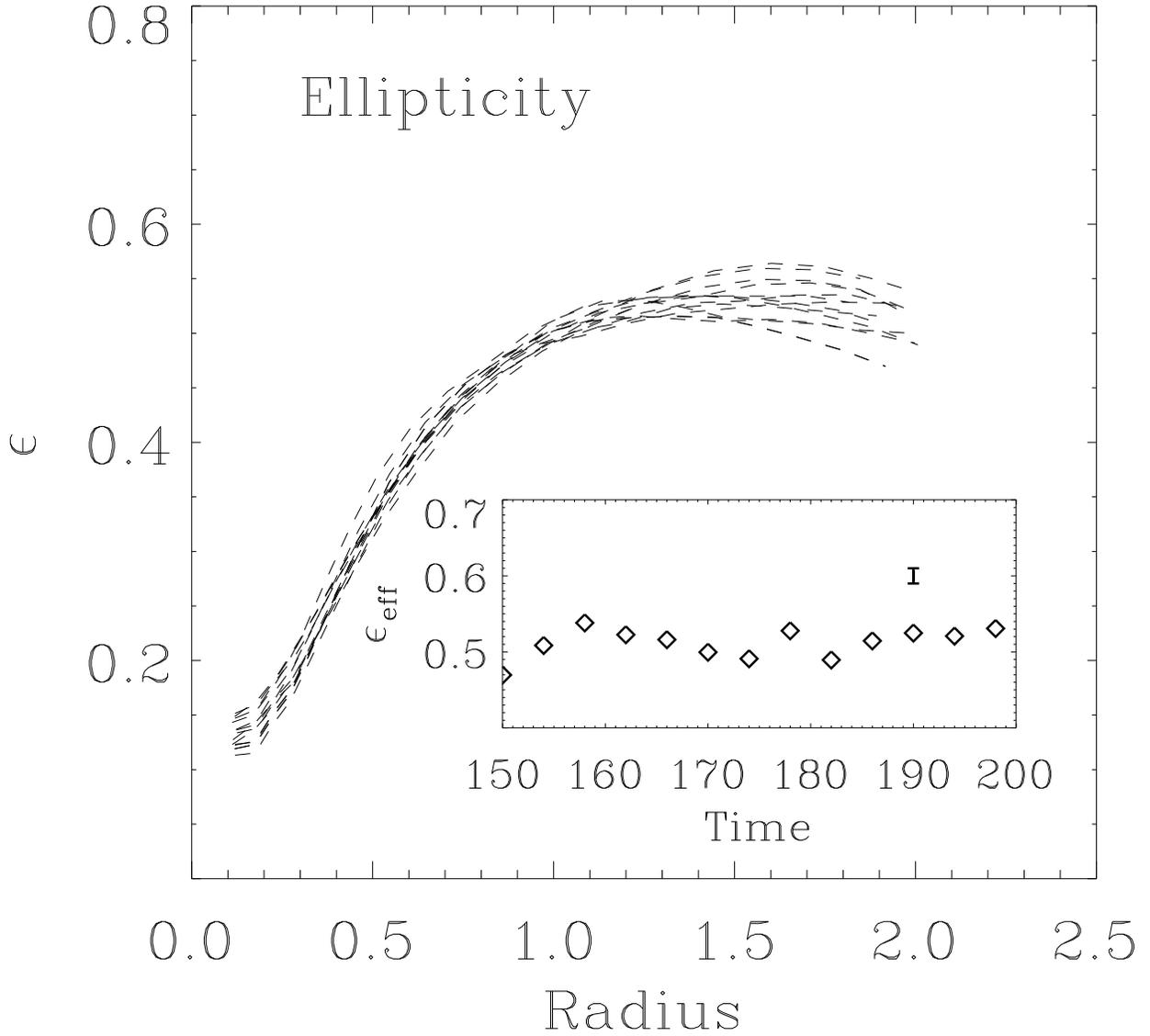}
\figcaption
{Radial change of the ellipticity $\epsilon$ at different evolutionary
times (dashed lines) and the time evolution of the characteristic
$\epsilon_{\mathrm{eff}}$ value for the 3:1 remnant number 10. The
error bar for $\epsilon_{\mathrm{eff}}$ was determined by
bootstrapping. \label{fig1}}
\end{figure}

The characteristic ellipticity $\epsilon_{\mathrm{eff}}$ for each
projection was defined as the isophotal ellipticity at $1.5
r_e$. Figure \ref{fig1} shows the ellipticity profiles for the 3:1
merger remnant 10 in intervals of four time units starting at
$t=150$. The ellipticity profile shows little evolution with time and
so does $\epsilon_{\mathrm{eff}}$ (small box). The behaviour of this
simulation is characteristic for the whole set of simulations. 

\placefigure{fig1}

\begin{figure}
\plotone{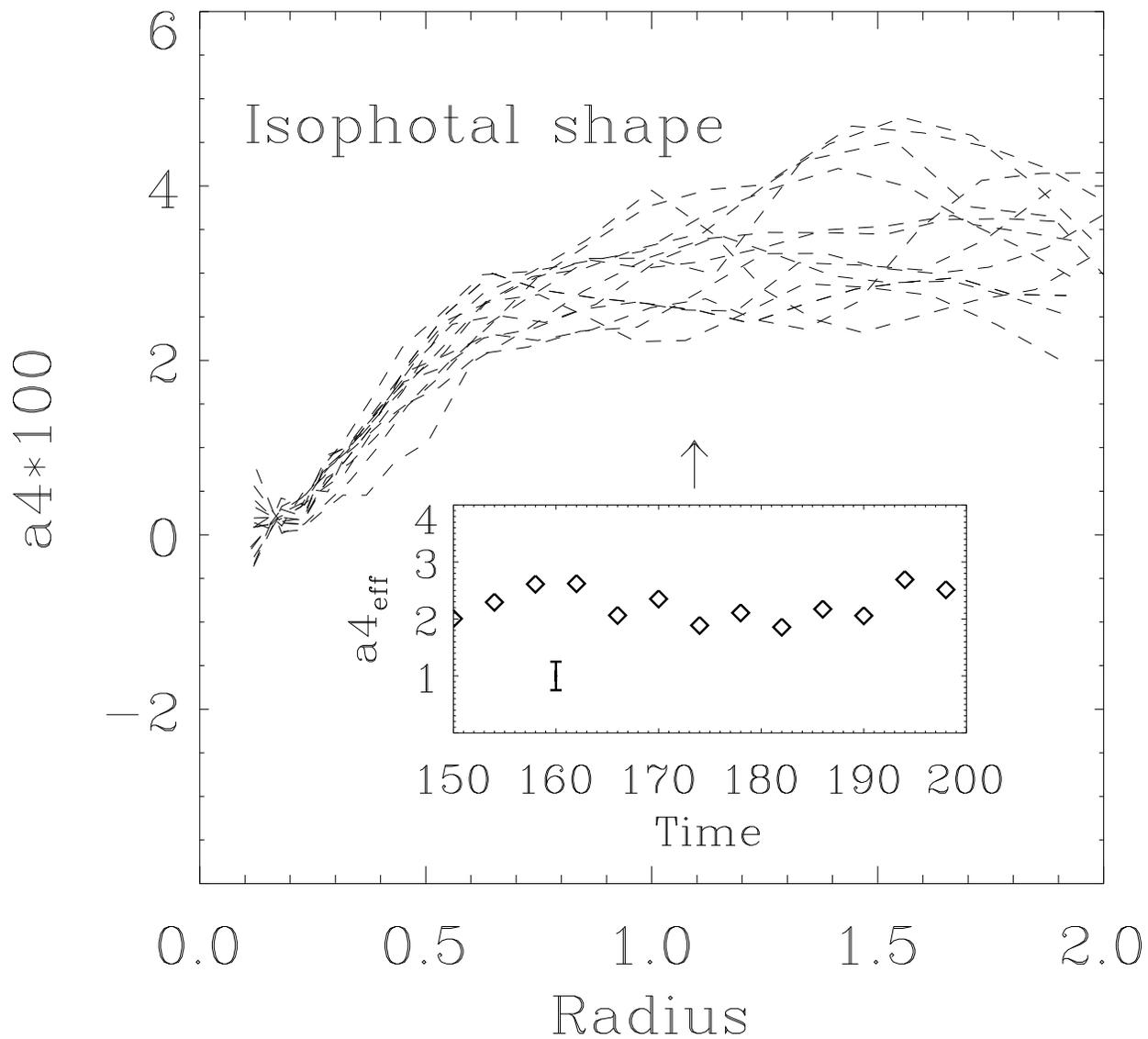}
\figcaption
{Dashed lines show the variation of $a_4$ along the apparent long axis
for different evolutionary times at intervals of $\Delta t =10$ for the 3:1 merger
remnant 10. The arrow indicates the value of the projected half mass
radius $r_e$. The small viewgraph shows the time evolution of the
effective $a4_{\mathrm{eff}}$ value assigned to each of the dashed
curves. The error bar was determined by statisitcal bootstrapping. \label{fig2}}
\end{figure}

\begin{figure}
\plotone{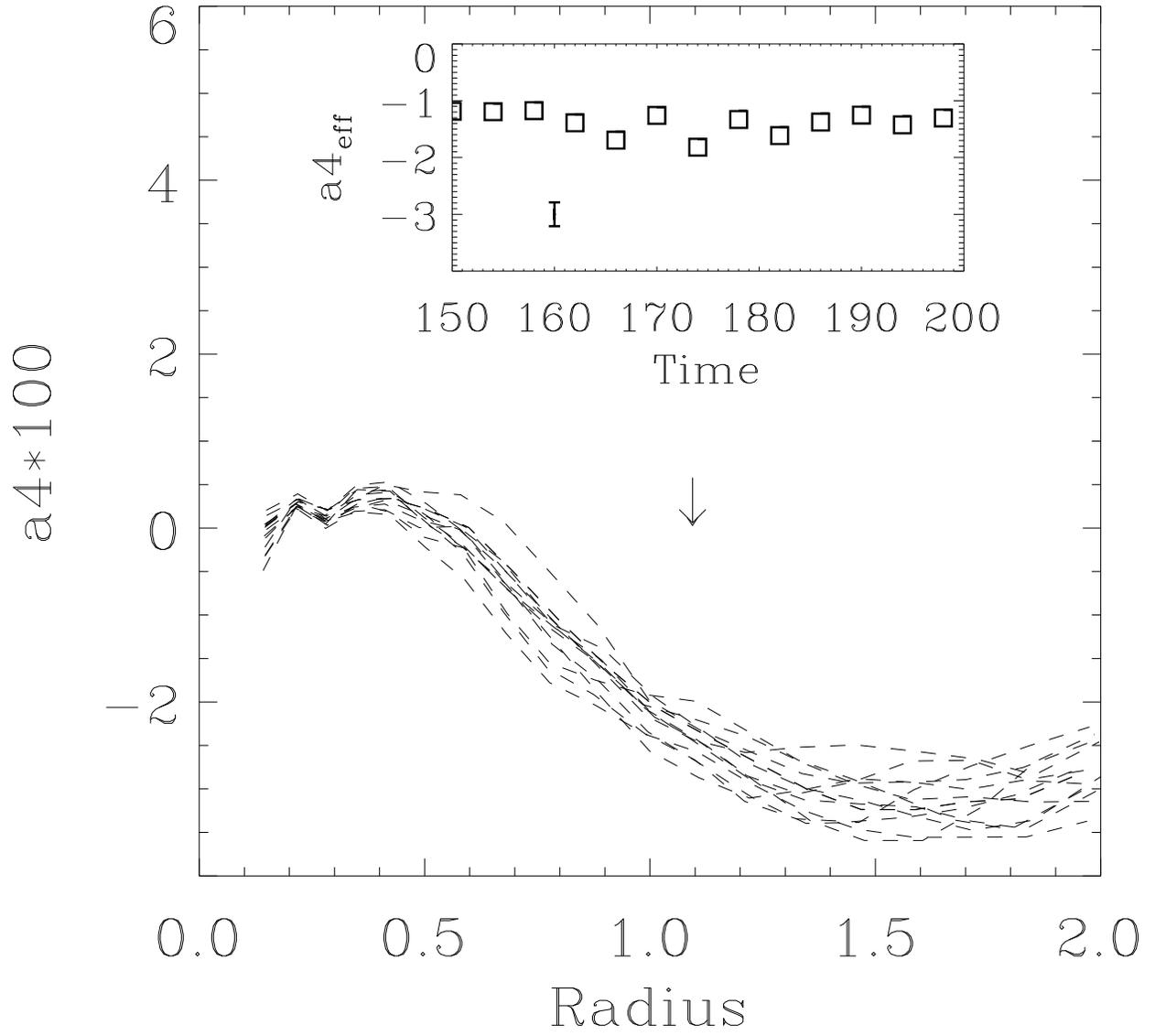}
\figcaption
{Same as Figure \ref{fig2} but for 1:1 merger remnant 2. \label{fig3}}
\end{figure}

Following the definition of BDM for the global properties of
observed giant elliptical galaxies, we determined for every
projection the effective $a_4$-coefficient, $a4_{\mathrm{eff}}$, as the
mean value of $a_4$ between $0.25 r_e$ and $1.0 r_e$, with $r_e$ being
the projected spherical half-mass radius. In case of a strong peak in
the $a_4$-distribution with an absolute value that is larger than the
absolute mean value, we chose the peak value. Being characteristic for
all merger remnants Figure \ref{fig2} shows the $a_4$-profiles at
different times for the 3:1 merger remnant with geometry 10. Figure
\ref{fig3} shows the corresponding $a_4$-profiles for the 
equal-mass remnant with geometry 2. Although the individual
$a_4$-profiles differ in details at different time steps the global 
characteristic shape is conserved. This is reflected in the time
evolution of $a4_{\mathrm{eff}}$ which
was assigned to each remnant at the different time steps as shown in
the small diagrams of Figure \ref{fig2} and Figure \ref{fig3}. The
change in $a4_{\mathrm{eff}}$ due to evolutionary effects (a real
change in the isophotal shape and/or a change in three-dimensional
shape leading to a different projection angle) is of the same order as
the error due to the limited particle number. The bootstrap error bar
is shown in the small diagrams of Figure \ref{fig1} and  Figure \ref{fig2}. 

\placefigure{fig1}

\placefigure{fig2}

These measurements convinced us that the photometric properties of the inner
regions (inside one effective radius) of the merger remnants evolve
relatively fast into an equilibrium configuration. As the
photometric properties do not significantly change with time they can
be used for a further statistical analysis.\\ 

To investigate projection effects we determined for each simulation
$a4_{\mathrm{eff}}$ and $\epsilon_{\mathrm{eff}}$ for 500 random
projections of every remnant. This resulted in 8000 projections for
$\eta=1$ and 16000 projections for $\eta=2,3,4$, respectively.

\begin{figure}
\plotone{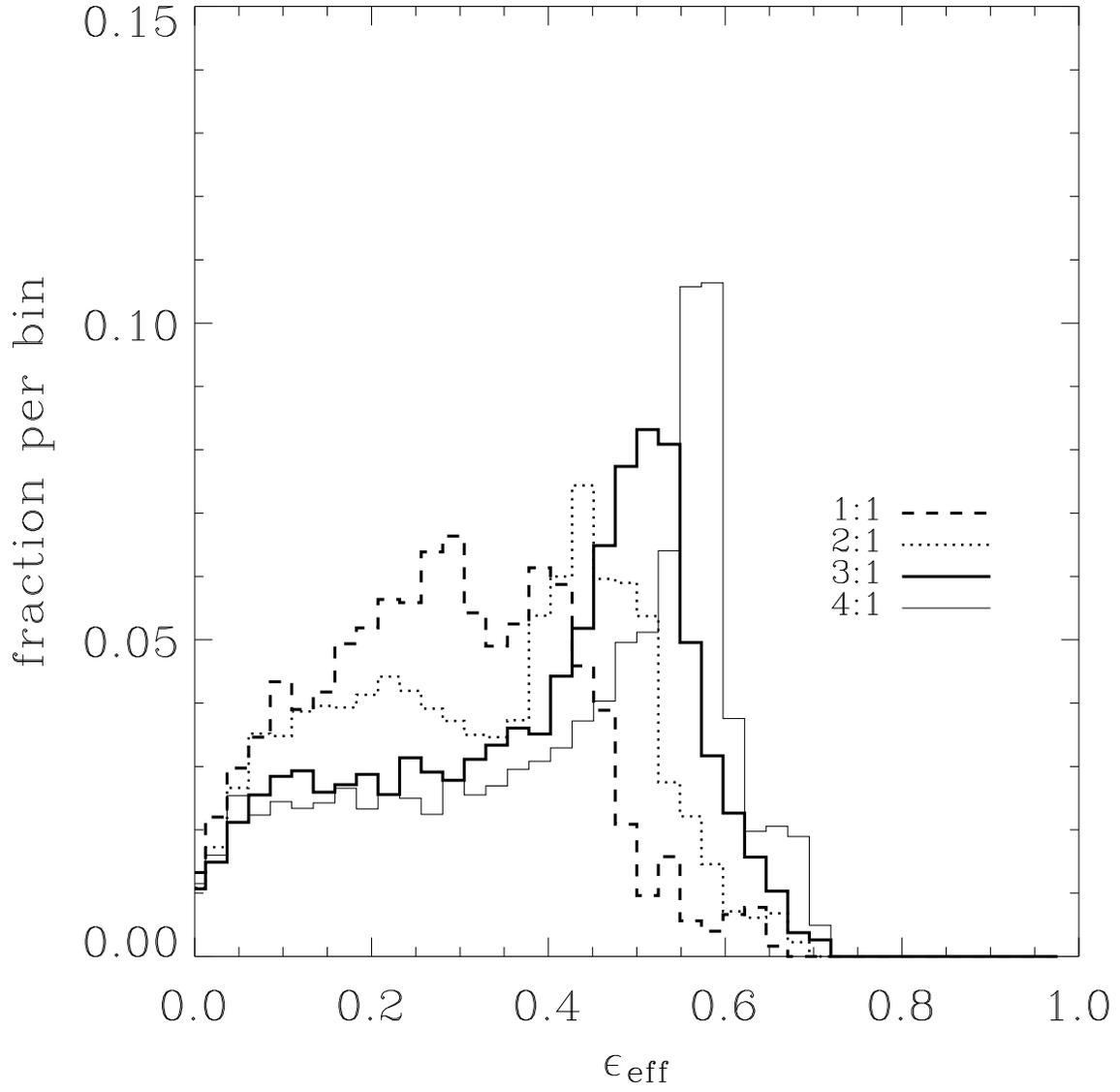}
\figcaption
{Normalized histograms of the effective ellipticity $\epsilon_{\mathrm{eff}}$ for
1:1, 2:1, 3:1, and 4:1 mergers.\label{fig4}}
\end{figure}

The normalized histograms of projected ellipticities are shown in
Figure \ref{fig4}. For equal-mass mergers the distribution rises from a
small number of projections with zero ellipticity to a peak
around  $\epsilon_{\mathrm{eff}} = 0.3$ and then falls to zero at
$\epsilon_{\mathrm{eff}} = 0.7$. 2:1  remnants show one peak around
$\epsilon_{\mathrm{eff}} = 0.25$ and a 
strong peak at $\epsilon_{\mathrm{eff}} = 0.45$ falling to zero at
$\epsilon_{\mathrm{eff}} = 0.7$. 3:1 and 4:1 remnants show a flat 
distribution for small ellipticities rising to a strong peak around
$\epsilon_{\mathrm{eff}} = 0.55$ and $\epsilon_{\mathrm{eff}} = 0.6$,
respectively. No projection leads to ellipticities larger than $0.7$. 

\begin{figure}
\plotone{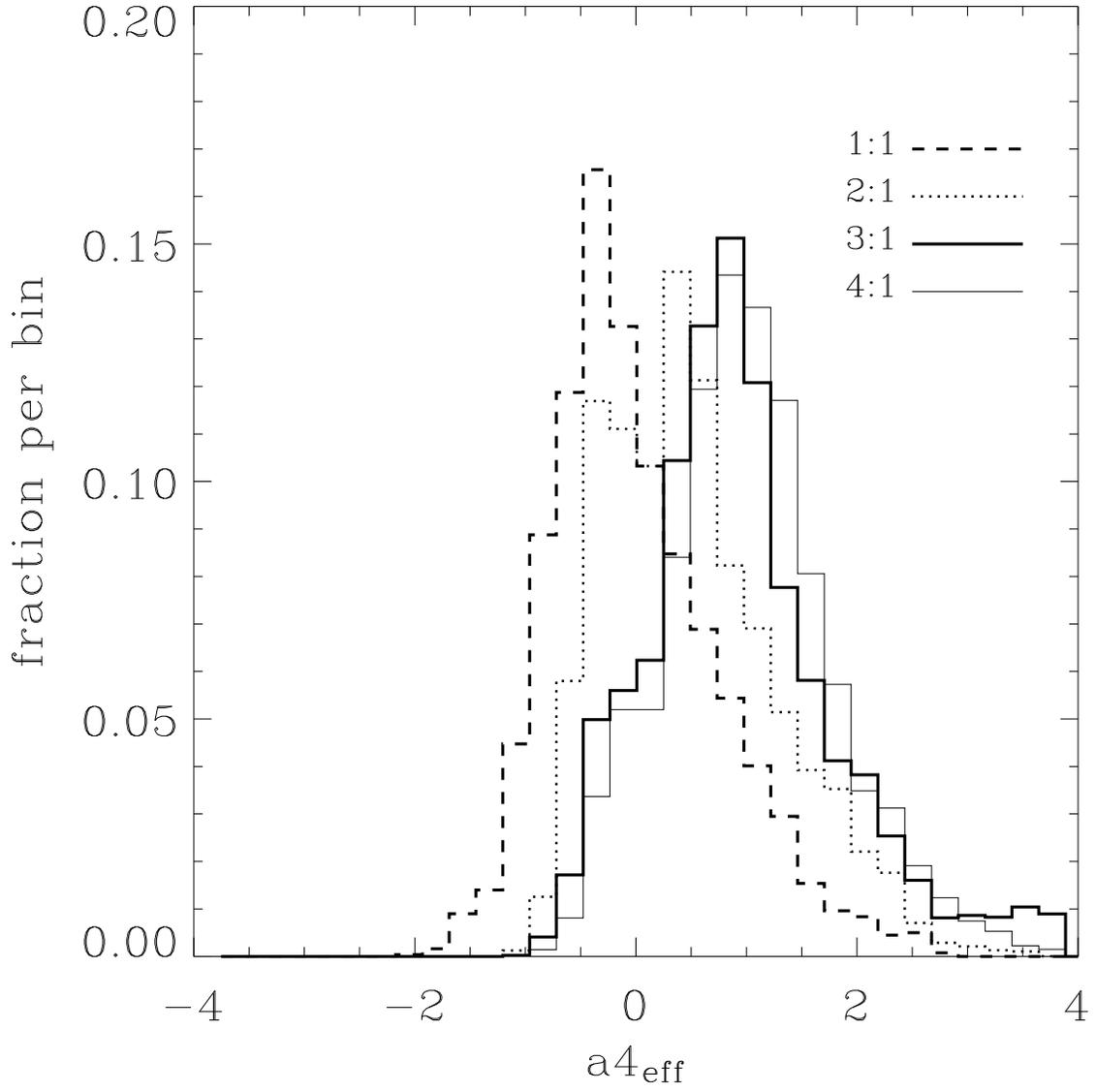}
\figcaption
{Normalized histograms of the shape parameter $a4_{\mathrm{eff}}$ for
1:1, 2:1, 3:1, and 4:1 mergers.\label{fig5}}
\end{figure}

The distribution function for the isophotal shapes of equal-mass
remnants peaks at $a4_{\mathrm{eff}} \approx -0.5$ and declines
rapidly for more negative values (Figure \ref{fig5}).  
In total 47\% of the projected 1:1 remnants show disky isophotes  
distributed in a broader wing for positive $a4_{\mathrm{eff}}$.
For higher mass ratios the distribution functions have a similar shape but
are shifted to more positive values of $a4_{\mathrm{eff}}$. 2:1
remnants peak at around $a4_{\mathrm{eff}} = 0.5$ and $73\%$  of the
projected remnants show disky isophotes.  The distributions for 3:1
and 4:1 remnants both peak around $a4_{\mathrm{eff}} \approx 1$. For
these systems the fraction of disky projections is about $89 \%$ and
$91 \%$, respectively.  

\placefigure{fig4}

\placefigure{fig5}

In addition, the 500 projected values for $a4_{\mathrm{eff}}$ and 
$\epsilon_{\mathrm{eff}}$ where used to calculate the two-dimensional     
probability density function for a given simulated remnant to be
"observed " in the $a4_{\mathrm{eff}}$-$\epsilon_{\mathrm{eff}}$
plane. We added up the probability densities assuming equal weights
for every simulation geometry at a given mass ratio. Figure
\ref{fig6} shows the results for mergers with  
$\eta = 1,2,3$, and $4$. The contours indicate the areas of 50\%
(dashed line), 70\% (thin line) and 90\% (thick line) probability to
detect a merger remnant with the given properties. Observed data
points from BDM are over-plotted. Filled boxes are boxy ellipticals with
$a4_{\mathrm{eff}} \le 0$ while open diamonds indicate disky ellipticals with
$a4_{\mathrm{eff}} > 0$. The errors were estimated applying the
statistical bootstrapping method \citep{hhs1994}. The area covered by
1:1 remnants with negative $a4_{\mathrm{eff}}$ is in good   
agreement with the observed data for boxy elliptical galaxies. In
particular the observed trend for more boxy galaxies to have 
higher ellipticities is reproduced. The disky equal-mass remnants also
follow the trend for observed disky ellipticals. 
Remnants with an $a4_{\mathrm{eff}}$ around zero can have slightly
higher ellipticities than observed. Remnants with $\eta = 2,3$, and $4$
predominantly populate the region of disky ellipticals and more disky
ellipticals also tend to be more flattened. There is a trend for
projections with $a4_{\mathrm{eff}} \approx 1$ to have slightly higher
ellipticities than observed (or at a given ellipticity the remnants are
not disky enough). Observed very disky ellipticals with relatively small
ellipticities can not be reproduced.   
 
\placefigure{fig6}

In summary there is a clear trend for unequal-mass mergers to 
produce more disky remnants. Responsible for the disky appearance of
the 3:1 and 4:1 remnants is the distribution of the particles of the
massive disk. Figure \ref{fig7} shows the different contributions from
the small and the large progenitor galaxy and the resulting isophotal
map of a characteristic 3:1 merger remnant. The particles originating
from the small progenitor accumulate in a torus-like structure that
has peanut-shaped or boxy isophotes. In contrast, the dominant
luminous material from the larger progenitor still keeps its disk-like
appearance. In combination, the contribution from the larger
progenitor -- since it is three times more massive -- dominates the
overall properties of the remnant. This result holds for all 3:1 and
4:1 merger remnants. The more massive disk component is not completely
destroyed during the merger event and determines the overall structure
of the remnant. For equal-mass mergers both disks are effected during
the merger and they can loose the memory of their initial state.

\placefigure{fig7}

The isophotal analysis also provides information about the radial 
change of the relative orientation of the major axes of the
isophotes. In general, the amount of isophotal twist depends 
on the projection angle. This is demonstrated in Figure \ref{fig8} for  
a more elongated and a nearly round projection of a 3:1 merger
remnant. To get a quantitative measure for the isophotal twist we
determined the relative position angle $\Delta \Phi$ between the
isophote at 0.5 $r_{\mathrm{eff}}$ and 1.5 $r_{\mathrm{eff}}$ for every
projection of the remnant. Figure \ref{fig9} shows a
comparison of isophotal twists for characteristic remnants with $\eta
=1,2,3$, and $4$. The isophotal twist is in general larger for
projections that appear nearly round. For ellipticities larger than
$\epsilon_{\mathrm{eff}} \approx 0.4$ the isophotal twist is $\Delta
\Phi \le 20^\circ$. The distribution of $\Delta \Phi$ versus
$\epsilon_{\mathrm{eff}}$ for random projections of every remnant is
consistent with observations of elliptical galaxies
\citep{bdm1988}. However, the simulated remnants can show larger
isophotal twists at high ellipticities than observed galaxies. In
addition there seems to exist no obvious correlation between the mass
ratio of the galaxies and the amount of isophotal twist.   

\placefigure{fig8}

\placefigure{fig9}

\subsection{Kinematics}

The central velocity dispersion $\sigma_0$ of every remnant was
determined as the average projected velocity dispersion of the 
luminous particles inside a projected galactocentric distance of $ 0.2
r_e $. We defined the characteristic rotational velocity along the
major and the minor axis as the projected rotational velocity at $1.5
r_e$ and $0.5 r_e$, respectively. Figure \ref{fig10} shows the time
evolution for these velocity measurements for a 3:1 merger simulation
which is characteristic for all simulations. The derived kinematical
properties of the remnants stay nearly constant over a long time
period and are therefore a good measurement of the intrinsic
kinematics of the simulated remnants. Again, the bootstrap error is of
the same order as the change of the measured values with time.

\placefigure{fig10}

Like for the isophotal shape we computed the statistical kinematical
properties of the simulated remnants and compared them with
observational data of elliptical galaxies. 

The normalized histograms for $(v_{\mathrm{maj}}/\sigma_0)$ are shown in Figure
\ref{fig11}.

\placefigure{fig11}

There  is a clear trend for 1:1 mergers to produce slowly rotating
ellipticals with $(v_{\mathrm{maj}}/\sigma_0)  < 0.3$. For 2:1 mergers the
peak value is around $(v_{\mathrm{maj}}/\sigma_0) = 0.4$. 3:1 and 4:1 merger
remnants have peaks around $(v_{\mathrm{maj}}/\sigma_0) = 0.6$ and $0.8$,
respectively. There are indications for a weak secondary peak around
$(v_{\mathrm{maj}}/\sigma_0) = 1$ for 3:1 and 4:1 remnants. 

Figure \ref{fig12} shows the distribution function in the
$(v_{\mathrm{maj}}/\sigma_0)$-$\epsilon_{\mathrm{eff}}$ plane. The area of
slowly rotating boxy ellipticals (filled boxes) is almost completely
covered by remnants of 1:1 mergers while 2:1 mergers have rotational
properties resembling faster rotating boxy and slowly rotating disky
ellipticals. 3:1 and 4:1 remnants are clearly fast rotating, show
high ellipticities and can be associated with fast rotating disky
ellipticals. Mergers with an increasing mass ratio produce faster
rotating ellipticals with higher ellipticities. All simulated data are
in good agreement with observations.

However, there is a trend for the simulated remnants to have slightly
larger maximum ellipticities than observed. Projections with high
ellipticities ($\epsilon > 0.5$) show systematically smaller values 
for $v_{\mathrm{maj}}/\sigma_0$ than observed. The effect is strongest for
equal-mass merger remnants and is discussed in detail below. This
indicates that the simulated systems are more strongly supported by
anisotropic velocity dispersions than the observed systems, despite
significant rotation for 3:1 and 4:1 remnants. Observed ellipticals
with $(v_{\mathrm{maj}}/\sigma_0) > 1.2$ can not be reproduced as was already
reported by \citet{cnr2001}.  

\placefigure{fig12}

The minor-axis kinematics of the simulated remnants was measured as the
rotation velocity along the minor axis at $0.5 r_{\mathrm{eff}}$. The amount of
minor axis rotation was parametrized as 
$(v_{min} / \sqrt{v_{\mathrm{maj}}^2 + v_{min}^2})$ \citep{b1985}. Minor axis
rotation in elliptical galaxies, in addition to isophotal twist, has
been suggested as a sign for a triaxial shape of the main body of
elliptical galaxies \citep{wbm1988,f1991}. Indeed, 1:1 mergers show
a significant amount of minor-axis rotation (Figure \ref{fig13}). This
is consistent with their mostly triaxial shape and the measured isophotal
twist. 3:1 and 4:1 remnants are more oblate and show weak minor axis 
rotation. However, the amount of isophotal twist is comparable to
equal-mass merger remnants. A detailed analysis of the connection
between intrinsic shape, isophotal shape,  and internal kinematics is
beyond the scope of this paper and needs further investigation.    

\placefigure{fig13} 

The anisotropy parameter  $(v_{\mathrm{maj}}/\sigma_0)^*$ was defined as the
ratio of the observed value of  $(v_{\mathrm{maj}}/\sigma_0)$ and the
theoretical value for an isotropic oblate rotator $(v/\sigma)_{theo} =
\sqrt{\epsilon_{\mathrm{obs}} / (1-\epsilon_{\mathrm{obs}})}$ with the
observed ellipticity $\epsilon_{\mathrm{obs}}$ \citep{b1978}. This
parameter has been used by observers as a test whether a given galaxy
with observed $v_{\mathrm{maj}}$, $\sigma_0$ and $\epsilon_{\mathrm{obs}}$ is
flattened by rotation ($(v_{\mathrm{maj}}/\sigma_0)^* \ge 0.7$)  or by velocity
anisotropy ($(v_{\mathrm{maj}}/\sigma_0)^* < 0.7$)
\citep{d1983,b1988,nch1988,sb1995}.

\placefigure{fig14}

Figure \ref{fig14} shows the normalized  histograms for the
$(v_{\mathrm{maj}}/\sigma_0)^*$ values of the simulated remnants. The 1:1
remnants peak around $(v_{\mathrm{maj}}/\sigma_0)^* \approx 0.3$ with a more
prominent tail towards lower values. Only about 10\% of the
projections have $(v_{\mathrm{maj}}/\sigma_0)^* > 0.6$. In contrast 82\% have
$(v_{\mathrm{maj}}/\sigma_0)^* < 0.4$ and would clearly be interpreted as  
being supported by anisotropic velocity dispersion. 2:1 remnants peak
at $(v_{\mathrm{maj}}/\sigma_0)^* \approx 0.5$ with 41\% showing
$(v_{\mathrm{maj}}/\sigma_0)^* > 0.6$.  3:1 and 4:1 mergers with peaks at
$(v_{\mathrm{maj}}/\sigma_0)^* \approx 0.6$ and $(v_{\mathrm{maj}}/\sigma_0)^*\approx
0.7$ are consistent with model predictions of oblate isotropic or slightly
prolate rotators. The percentage of projections with
$(v_{\mathrm{maj}}/\sigma_0)^* > 0.6$ is 64\% and 83\%, respectively. 3:1 and
4:1 remnants also have predominantly disky isophotes and 
cover the area populated by observed disky ellipticals in the $log
(v_{\mathrm{maj}}/\sigma_0)^*$ - $a4_{\mathrm{eff}}$ diagram (Figure
\ref{fig15}).     

1:1 merger remnants show the most complex behaviour. A significant
number of the projected remnants lie in a region of disky anisotropic
systems where no elliptical galaxy is observed. Assuming that a
projected remnant clearly fails to resemble an observed elliptical
if it has  $a_4 > 0.3$ and $(v_{\mathrm{maj}}/\sigma_0)^* < 0.5$, 28\% of the
1:1 remnants would fall into the forbidden regime. \\

\placefigure{fig15}

To understand the behaviour of equal-mass remnants in more
detail we illustrate the influence of the initial orientation of the
progenitor disks on the properties of the merger remnants. We defined
mean values of the investigated quantities of 
all 500 projections for every merger geometry. Thereafter we divided
the equal-mass remnants into four groups of geometries that produce
remnants with almost distinct properties: slowly rotating  boxy
remnants with $\overline{v_{\mathrm{maj}}/\sigma_0} < 0.2$ and $\overline{a_4}
< -0.4$ (group A; geometries 2,4,5, and 14), slowly rotating remnants
with predominantly disky isophotes having $\overline{v_{\mathrm{maj}}/\sigma_0}
< 0.2$ and $\overline{a_4} > 0.25$ (group B; geometries 1,8,9,11, and 
13), modestly rotating remnants ($\overline{v_{\mathrm{maj}}/\sigma_0} > 0.2$)
that appear to be isotropic with $\overline{(v_{\mathrm{maj}}/\sigma_0)^*} >
0.7$ (group C; geometries 7, 12, and 15) and remnants with modest
deviations in isophotal shape $-0.2 < \overline{a_4} < 0.2$ (group D;
geometries 3,6,12, and 16). The results are shown in Figure \ref{fig16}. 

\placefigure{fig16}

Remnants of group A are slowly rotating anisotropic systems with
purely boxy isophotes. More than 90\% of the projected remnants are
consistent with observations of boxy anisotropic elliptical
galaxies. This class of equal-mass merger remnants was  
sampled by \citet{nbh1999}. Most projections of group B show disky
isophotes and appear to be anisotropic. Around 60\% of the projected
remnants of this group fail to fall on the observed
correlation. Furthermore the remnants of this group show too high 
ellipticities at low rotation velocities, a behaviour already
described by \citet{hhs1996}. The shaded regions in Figure
\ref{fig16} indicate the location of the highly elliptical projections
($\epsilon > 0.4$) . They dominate the disky anisotropic branch of
"not observed" merger remnants. It is possible that the very elongated
projections of group B would be classified as S0
galaxies. Some of them, as NGC4550, show weak net rotation and are
still very elongated. Geometry 1 produces a very axisymmetric remnant with two
counterrotating populations of stars in as it is observed
in NGC4550 (see e.g. \citet{pfe1999}). However, \citet{rix1992} argued
that a  merger origin for this particular galaxy is unlikely as it
would have heated both disks too much. 

Group C combines all rotating 1:1 remnants. They appear slightly disky
or boxy depending on the orientation and are isotropic. It has to be
noted that even the disky remnants of this group are consistent with 
observations and only 10\% of the projections are failures. This group
shows a clear connection  to the initial conditions as the spins of
the progenitor disks were almost aligned. Projections of 
group D can either have disky or boxy isophotes and are mostly
anisotropic. 35\% of the projections fall in the forbidden regime.

\section{Conclusions \& Discussion}
\label{con}
We used a large set of N-body simulations of collisionless mergers of
disk galaxies with mass ratios of $\eta =1,2,3,$ and $4$ to 
investigate for the first time statistical properties of disk-merger
remnants. In contrast to previous studies where only projections
along the principal axes were investigated we analyzed a large
number of randomly projected merger remnants and compared the results
in a consistent way to observations. Galaxies on the sky are also
viewed along random lines of sight therefore this is the appropriate
way of comparison.    

We showed that the detailed isophotal and kinematical
properties of the simulated merger remnants reach their equilibrium
values relatively fast after the merger of the central parts of the
galaxies is complete and thereafter do not significantly evolve with
time. The high resolution of the simulations made it possible to keep
the errors, especially for $a_4$, at a reasonably low value and
enabled us to perform a statistical analysis.  

The basic result is that the projected kinematical and photometric
properties of remnants of major mergers of disk 
galaxies are in surprisingly good agreement with the observational data
for elliptical galaxies. The mass ratio of the progenitor disks
determines the global properties of the remnants. The influence of the
initial orientation on the remnants is strongest for equal-mass
mergers.

Purely boxy, anisotropic and slowly rotating remnants with a large
amount of minor axis rotation can only be produced by equal-mass
mergers with certain initial orientations. However, if the 
initial spins of the disks were almost aligned the remnants appear to be
isotropic and disky or boxy, depending on the viewing angle. In total
28\% of all projected 1:1 remnants show properties that are
not observed at all. As some of these remnants have high
ellipticities and very small $(v_{\mathrm{maj}}/\sigma_0)^*$ they appear to 
be flattened by strongly anisotropic velocity dispersions. In
addition they mostly show very disky isophotes. As we can not
exclude that mergers with these geometries have occurred they should
be observed, which is not the case.  If this
controversy is not solved in the future it might constitute a serious
problem for the merger hypothesis.  However, if observed, these
elongated disky objects might have been classified as S0s and have
therefore been excluded from the observed sample of bona fide ellipticals.  

Based on the diverse properties of the 1:1 remnants 
they more likely resemble the class of intermediate mass giant
ellipticals in the transition region from rotating disky to non-rotating
boxy ellipticals. In this regime even boxy rotating galaxies have been
observed \citep{nb1989}.     

In contrast, 3:1 and 4:1 mergers form a more homogeneous group of
remnants. They have preferentially disky isophotes, are fast rotating
and show small minor axis rotation independent of the assumed
projection. In general the properties are in very good agreement with
observations of fast rotating disky ellipticals.  

2:1 mergers have intermediate properties with boxy or disky isophotes
depending on the projection and the orbital geometry of the
merger. Globally, different projection angles and orbits do not change
the fundamental properties of 1:1 mergers on the one side and 3:1 and
4:1 mergers on the other side.\\ 

In summary, many photometric and kinematical properties of low and
intermediate mass giant elliptical galaxies can be understood as a
sequence of major mergers of disk galaxies with varying mass
ratios. The homogeneous group of massive boxy ellipticals populates the
boxy anisotropic area of the $(v_{\mathrm{\mathrm{maj}}}/\sigma_0)^*$ - $a4_{\mathrm{eff}}$
plane. Their distribution is inconsistent with the distribution of
simulated 1:1 remnants.  Therefore they are most likely  not remnants of
equal-mass mergers of disk galaxies. Boxy and mildly anisotropic remnants
are only reproduced by a small subsample of initial conditions. There
is no reason why other geometries should have been avoided. It is more
likely that massive boxy ellipticals have formed by other processes
like mergers of early type galaxies \citep{nb2000} or multiple
mergers in a group environment \citep{wh1996}.\\  

The simulations presented here were purely dissipationless, taking into
account only the stellar and dark matter component of a galaxy. The
importance of gas in galaxy-galaxy mergers and the detailed influence
on the structure of elliptical galaxies is not fully understood up to
now. Numerical simulations of galaxy mergers including gas by
\citet{bh1996} and \citet{b1998} have shown that the presence of gas can change
the orbital structure and the shape of the merger remnants. As soon as
the gas is driven to the center during the merger the mass
concentration seems to be responsible for the destruction of stellar
box orbits. This process makes it even more difficult to explain the
formation of giant boxy ellipticals by binary disk mergers. \\

\begin{figure}
\plotone{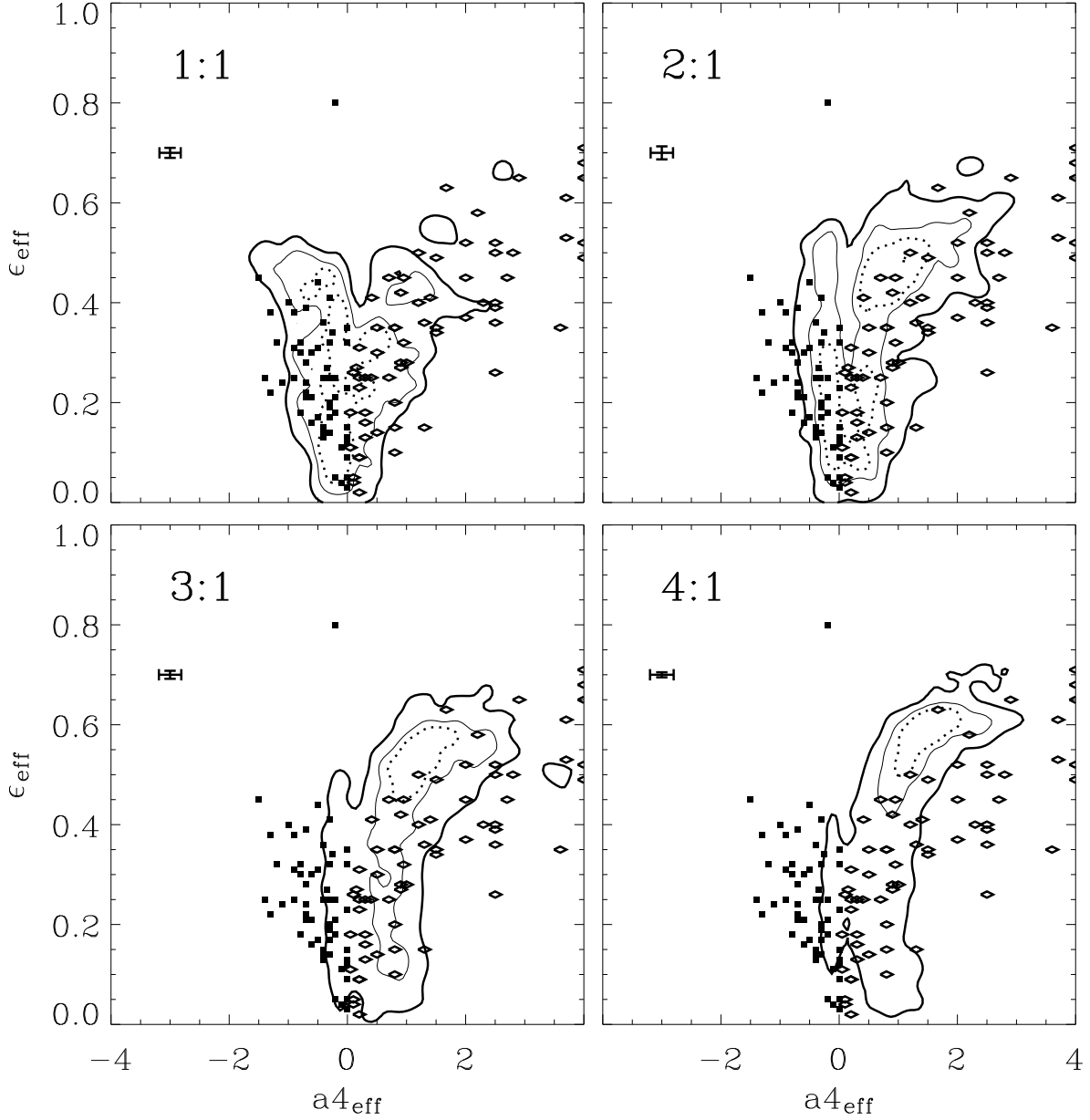}
\figcaption
{Ellipticities $\epsilon_{ell}$  versus the characteristic shape parameter
$a4_{\mathrm{eff}}$ for 1:1, 2:1, 3:1, and 4:1 mergers. The
contours indicate the 50\% (dotted line), the 70\% (thin line) and the
90\% (thick line) probability to find a merger remnant in the enclosed area.
The error bars for $a4_{\mathrm{eff}}$ are derived applying statistical
bootstrapping. Errors for $\epsilon_{\mathrm{eff}}$ are too small to be visible on this plot.
Black boxes indicate values for observed boxy elliptical galaxies, open
diamonds show observed disky ellipticals (data from BDM).\label{fig6}}
\end{figure}

\begin{figure}
\epsscale{0.5}
\plotone{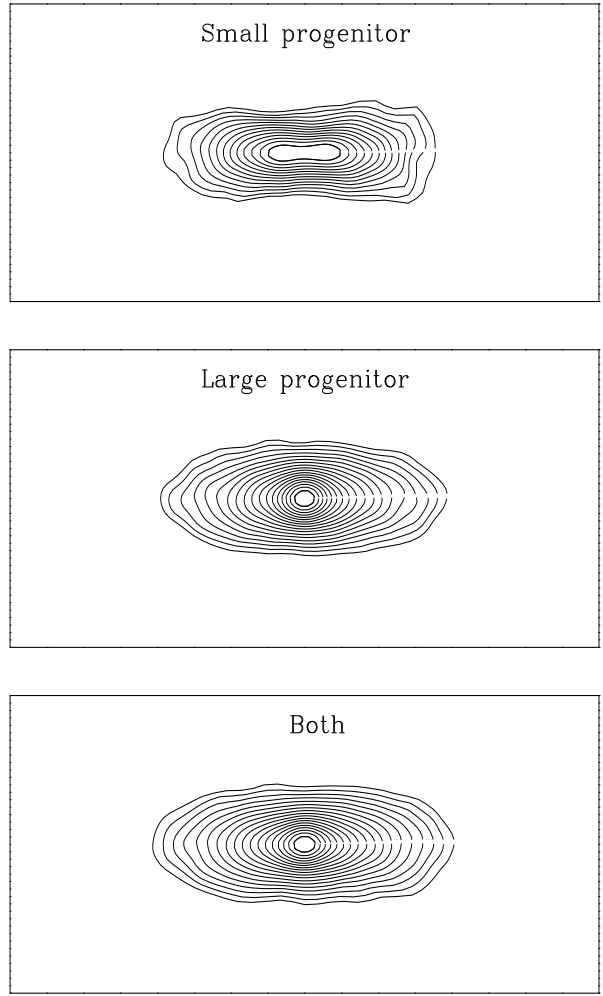}
\figcaption
{Characteristic isodensity contours of a typical 3:1 merger remnant. The
two upper panels show the contours for the luminous particles of
the smaller and larger progenitor, separately. The lower panel shows the
resulting contours of the complete remnant. \label{fig7}}
\end{figure}

\begin{figure}
\epsscale{1}
\plotone{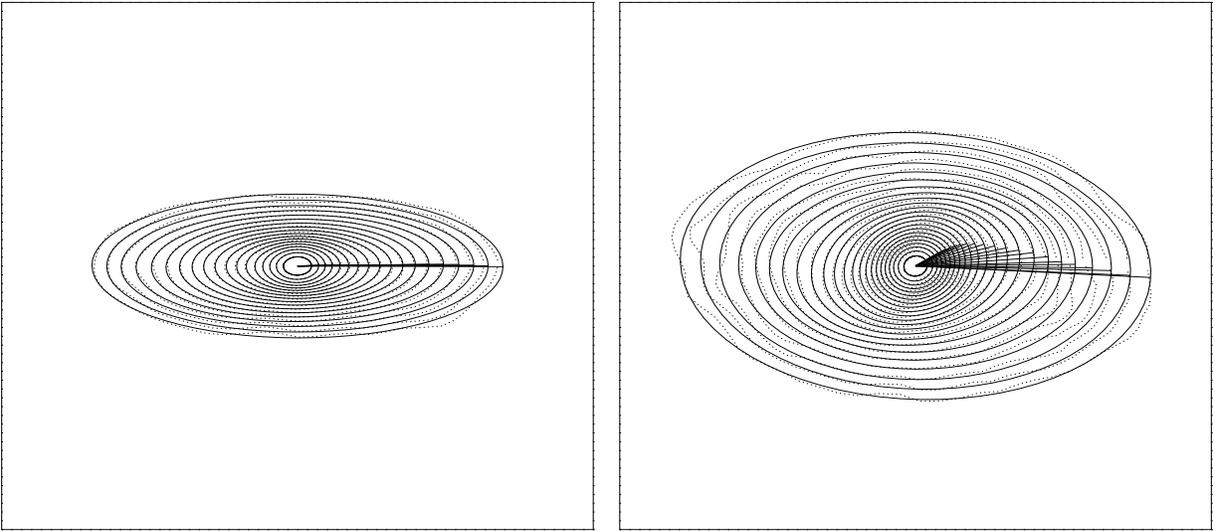}
\figcaption
{The real isophotes (dotted) and the best fitting ellipses (solid
line) a typical 3:1 merger remnant seen in a more elongated (left) and
rounder (right) projection. The major axes of the fitted ellipses are
plotted for every isodensity contour. The change of direction of the
axes indicates the isophotal twist. The box length is 3 length
units.\label{fig8} } 
\end{figure}

\begin{figure}
\plotone{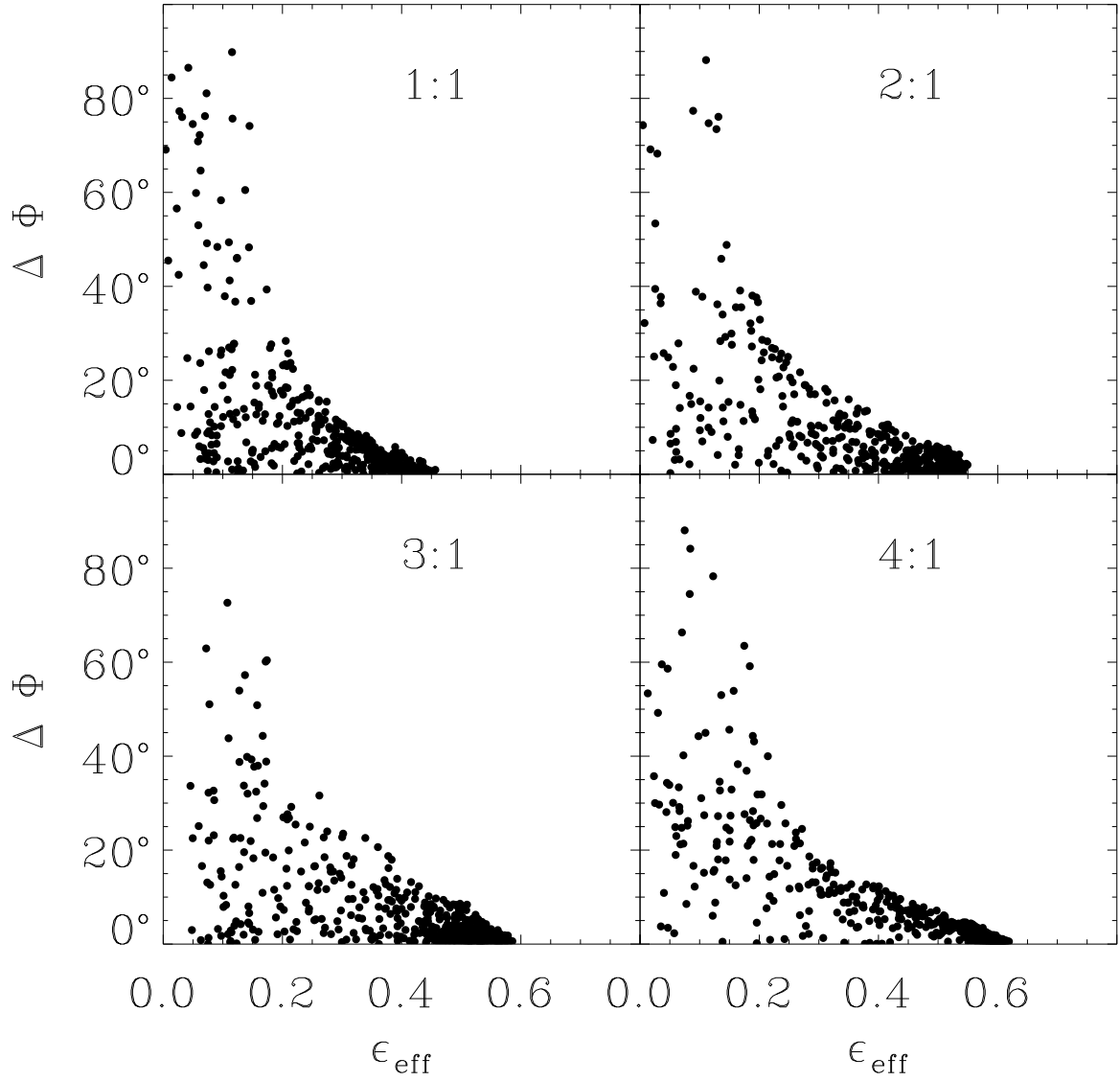}
\figcaption
{Isophotal twists $\Delta \Phi$ vs. ellipticity $\epsilon_{\mathrm{eff}}$ for a 
characteristic 1:1, 2:1, 3:1, and 4:1 merger remnant, respectively. For every mass ratio
500 random projections are shown. \label{fig9}}
\end{figure}

\begin{figure}
\plotone{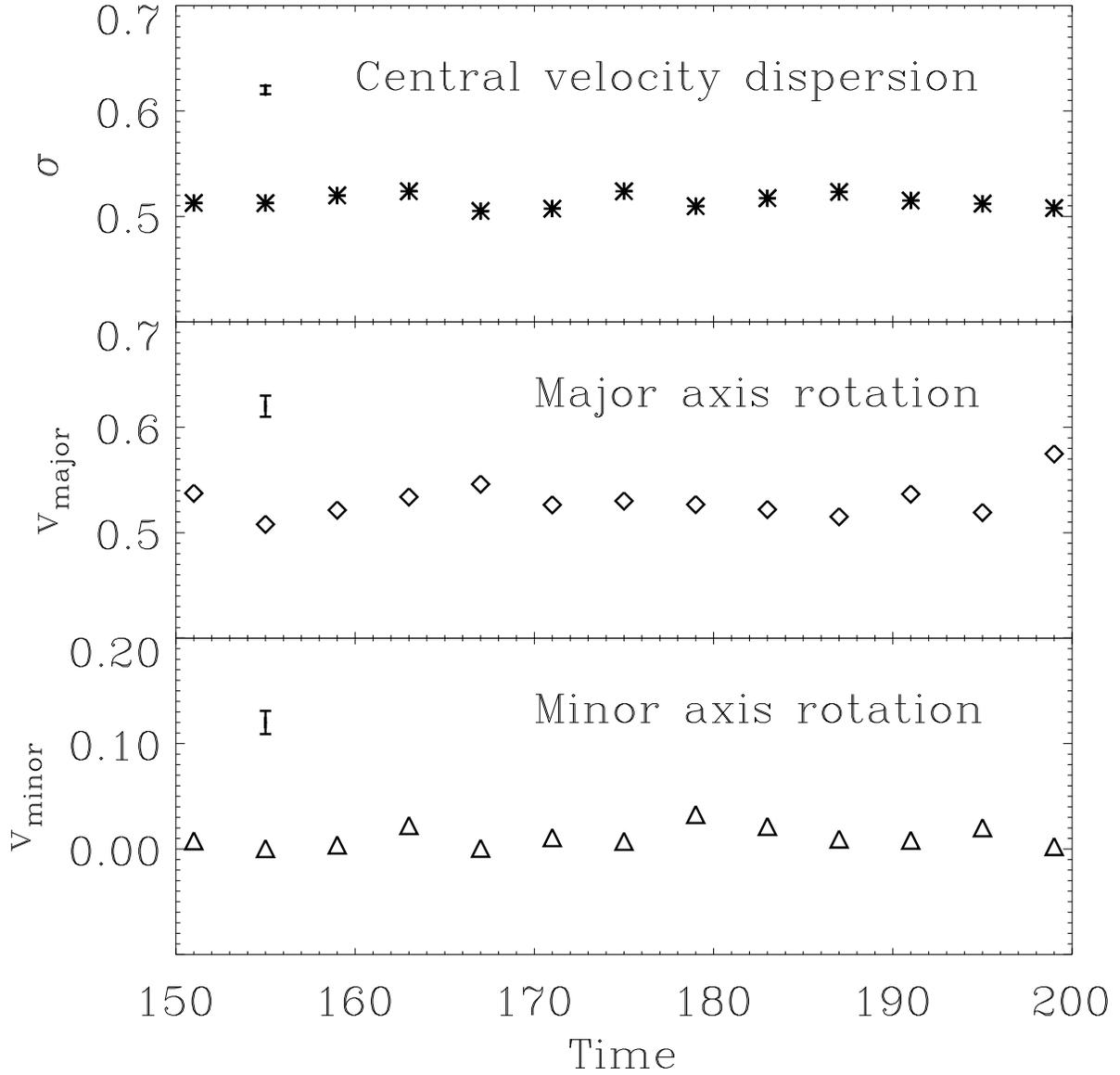}
\figcaption
{Time evolution of the characteristic kinematical parameters for a 3:1
  remnant: projected central velocity dispersion $\sigma$, major axis
  rotation velocity $v_{\mathrm{major}}$ at 1.5 $r_e$, and minor axis rotation
  velocity $v_{\mathrm{minor}}$ at 0.5 $r_e$. The error bars indicate the
  bootstrap error. \label{fig10}} 
\end{figure}

\begin{figure}
\plotone{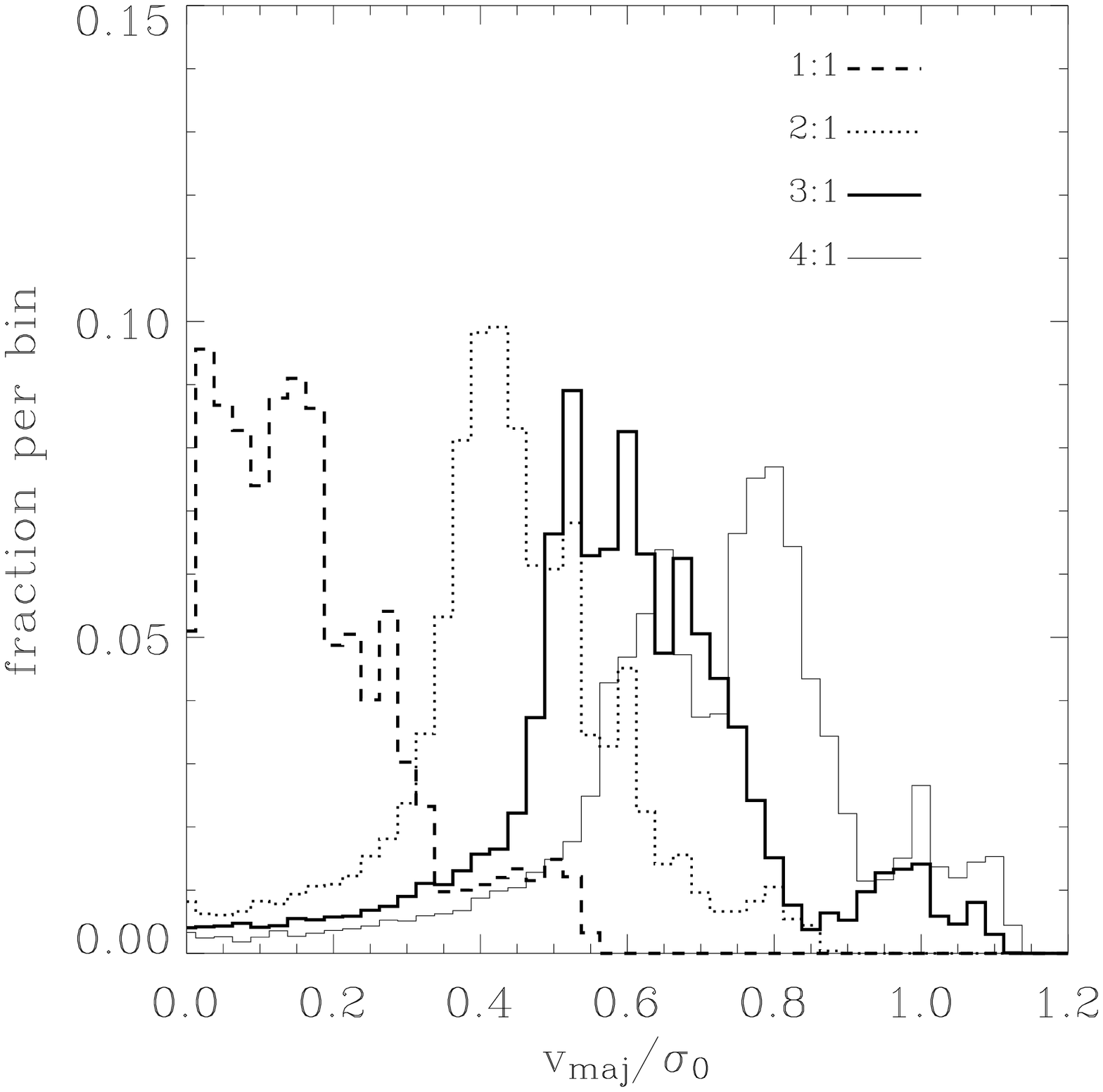}
\figcaption
{Normalized histograms of $v_{\mathrm{maj}}/\sigma_0$ for 1:1, 2:1, 3:1, and
4:1 mergers.\label{fig11}}
\end{figure}

\begin{figure}
\plotone{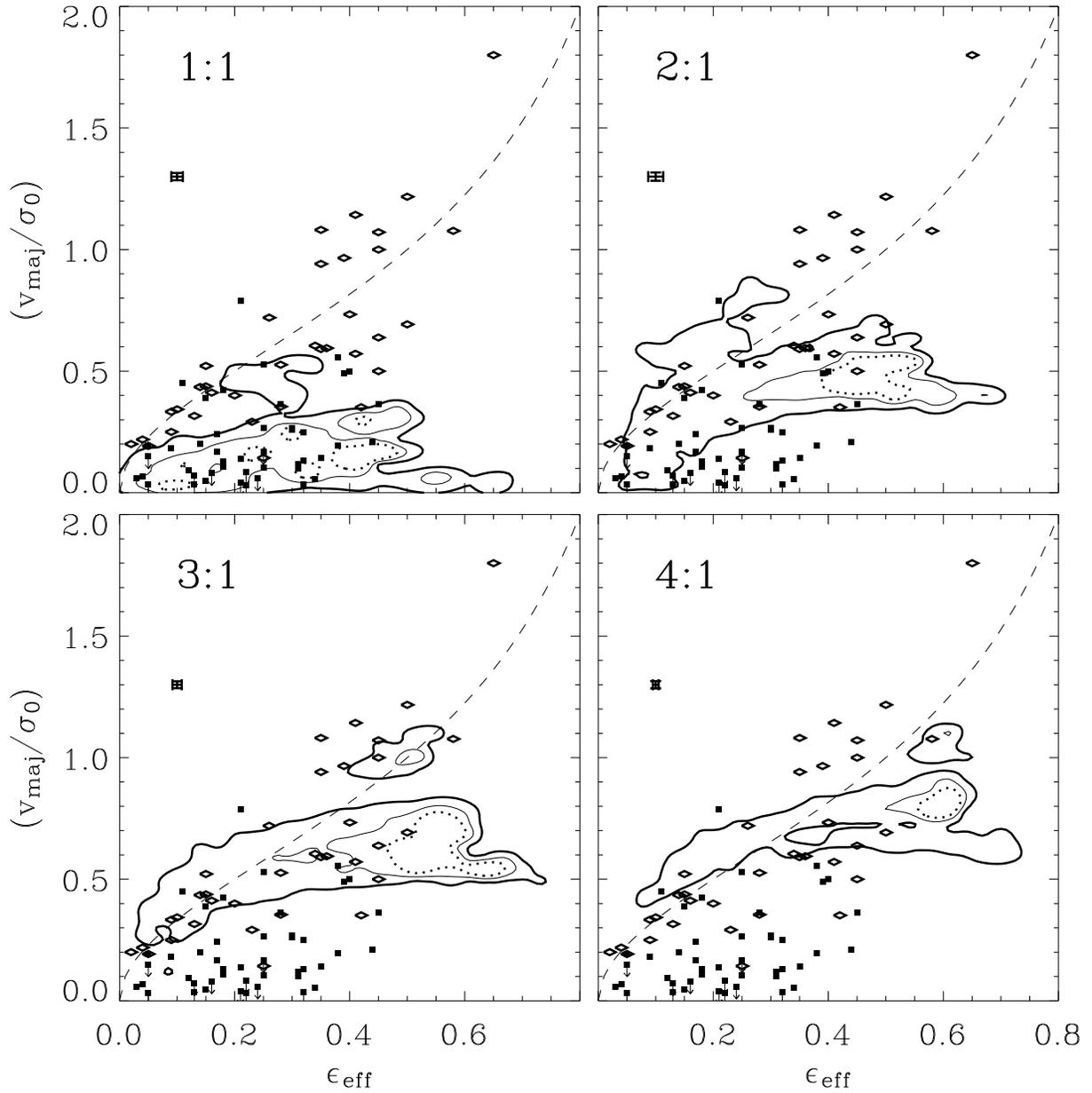}
\figcaption
{Rotational velocity along the major axis $v_{\mathrm{maj}}$ over the central
velocity dispersion $\sigma_0$ versus the characteristic ellipticity
$\epsilon_{\mathrm{eff}}$ for 1:1, 2:1, 3:1 and 4:1 
mergers. The contours indicate the 50\% (dotted line), the 70\% (thin line) and the
90\% (thick line) probability to find a merger remnant in the enclosed area. 
Values for observed ellipticals are overplotted. Filled boxes
indicate data for boxy elliptical galaxies, open diamonds show data
for disky ellipticals. Arrows indicate upper limits. The dashed line
shows the theoretical value for an oblate isotropic rotator. The error
bars were derived by statistical 
bootstrapping.\label{fig12}}
\end{figure}

\begin{figure}
\plotone{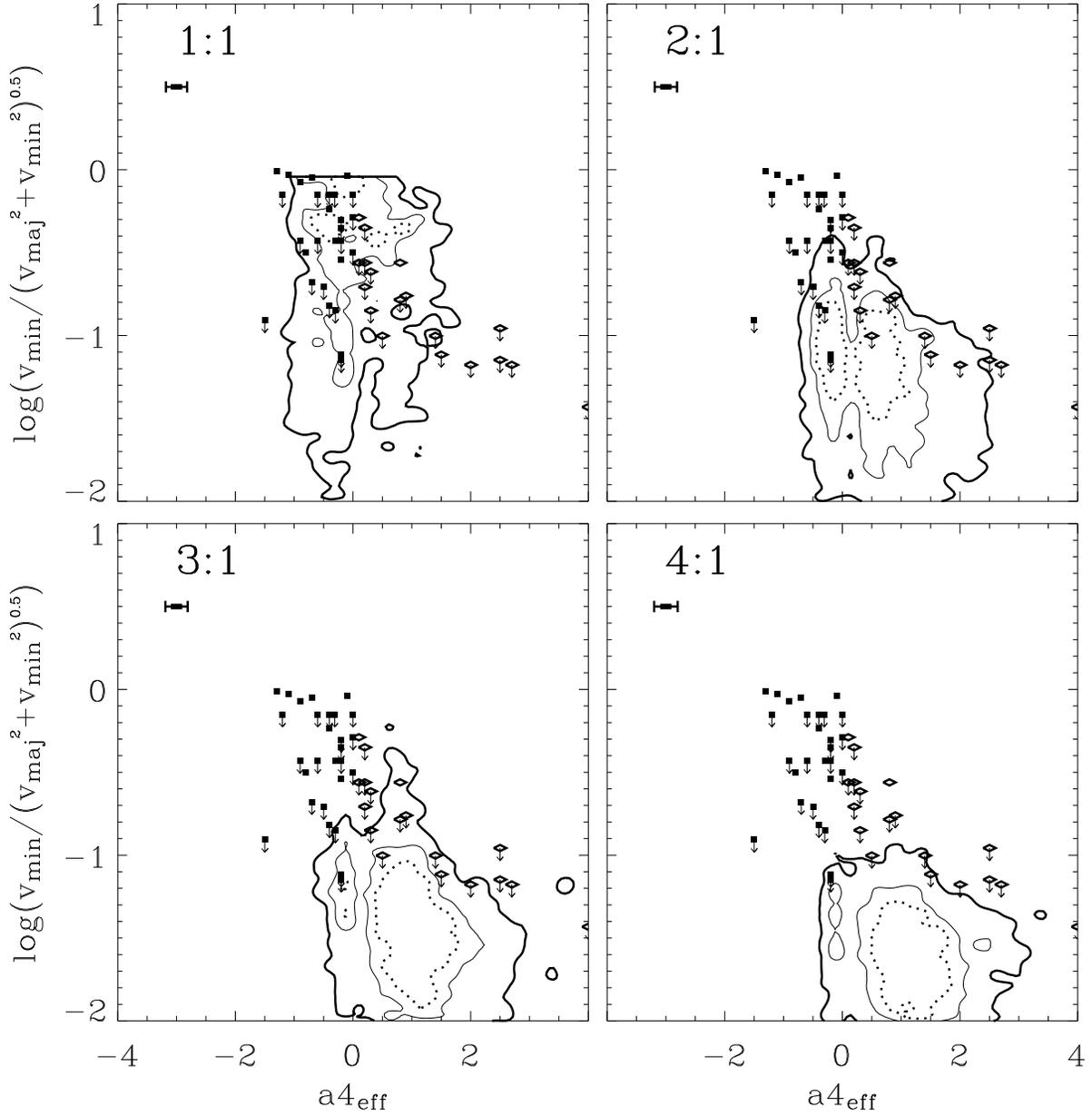}
\figcaption
{Amount of minor axis rotation $log(v_{\mathrm{min}} / \sqrt{v_{\mathrm{maj}}^2 +
v_{\mathrm{min}}^2})$ versus  $a4_{\mathrm{eff}}$ for 1:1, 2:1, 3:1 and 4:1
mergers.The
contours indicate the 50\% (dotted line), the 70\% (thin line) and the
90\% (thick line) probability to find a merger remnant in the enclosed area.
 Values for observed ellipticals are
overplotted. Filled boxes indicate data for boxy elliptical
galaxies, open diamonds show data for disky ellipticals. The data with 
arrows provide an upper limit. The error 
bars were derived by statistical 
bootstrapping.\label{fig13}}
\end{figure}

\begin{figure}
\plotone{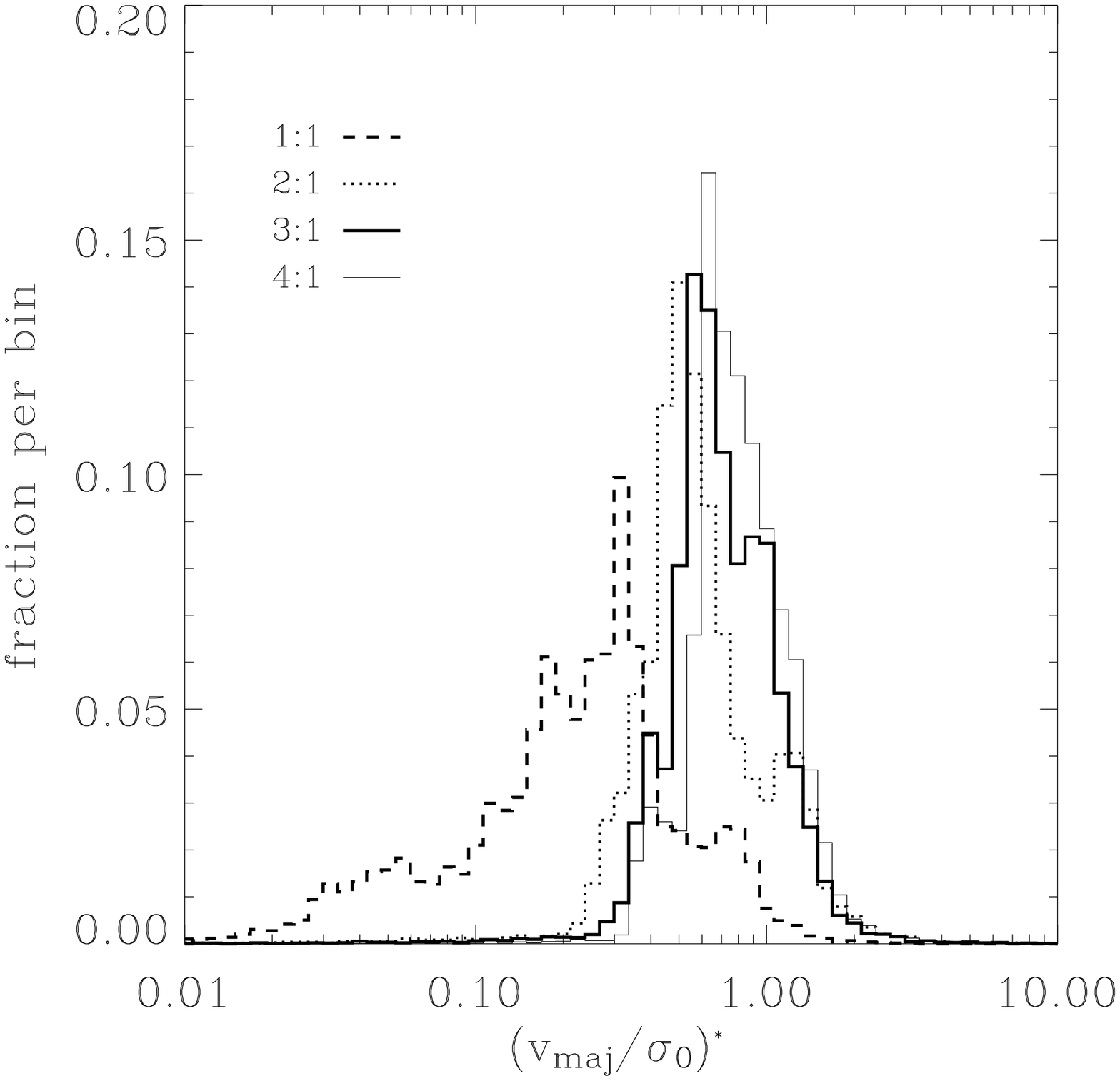}
\figcaption
{Normalized histograms of $(v_{\mathrm{maj}}/\sigma_0)^*$ for 1:1, 2:1, 3:1, and
4:1 mergers.\label{fig14}}
\end{figure}

\begin{figure}
\plotone{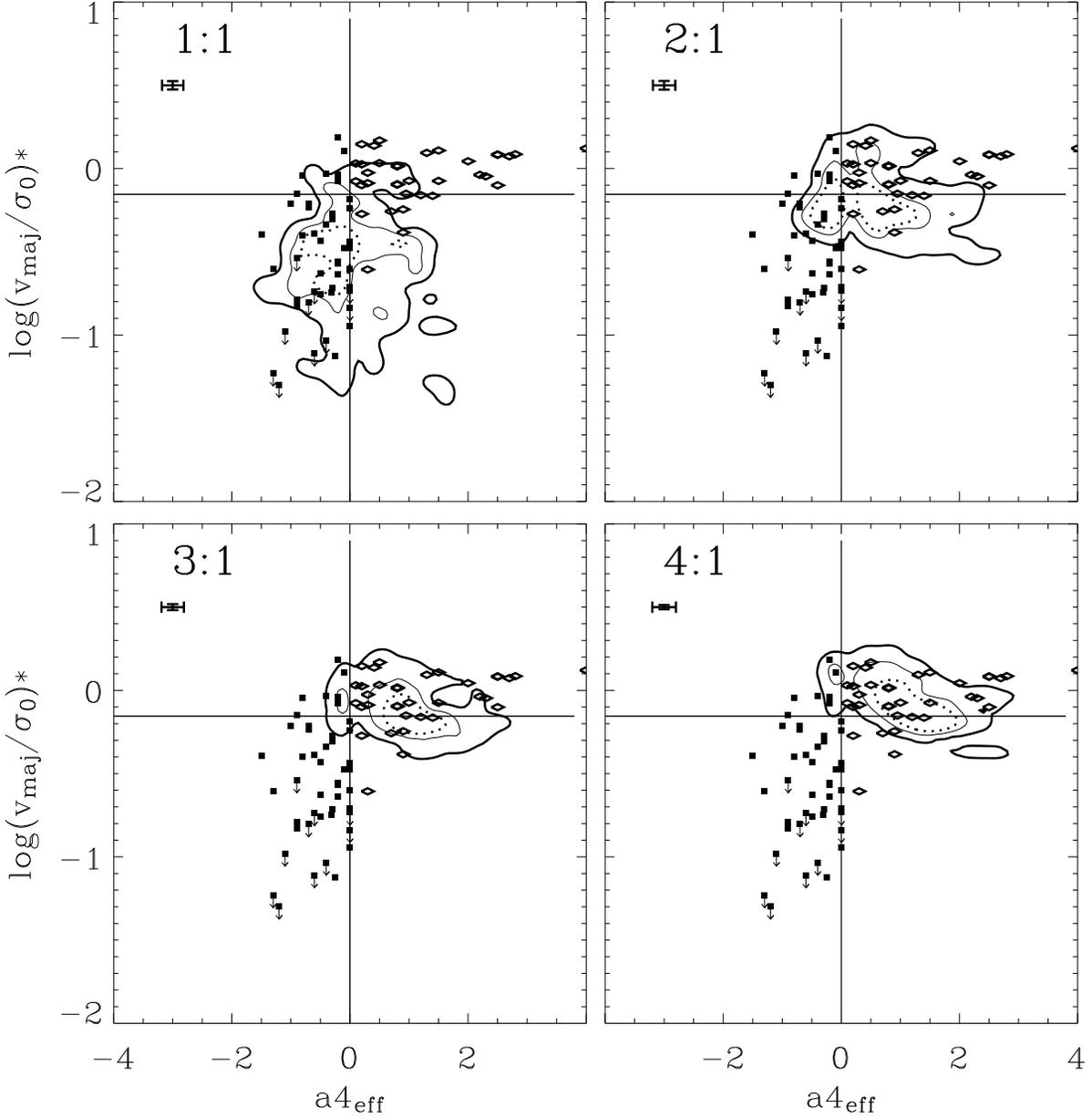}
\figcaption
{Anisotropy parameter $(v_{\mathrm{maj}}/\sigma_0)^*$  versus  $a4_{\mathrm{eff}}$ for
1:1, 2:1, 3:1 and 4:1 mergers.The
contours indicate the 50\% (dotted line), the 70\% (thin line) and the
90\% (thick line) probability to find a merger remnant in the enclosed area. 
Values for observed
ellipticals are overplotted. Filled boxes indicate data for boxy elliptical
galaxies, open diamonds show data for disky ellipticals. Arrows
indicate upper limits. The error bars 
were derived by statistical
bootstrapping.\label{fig15}}
\end{figure}

\begin{figure}
\plotone{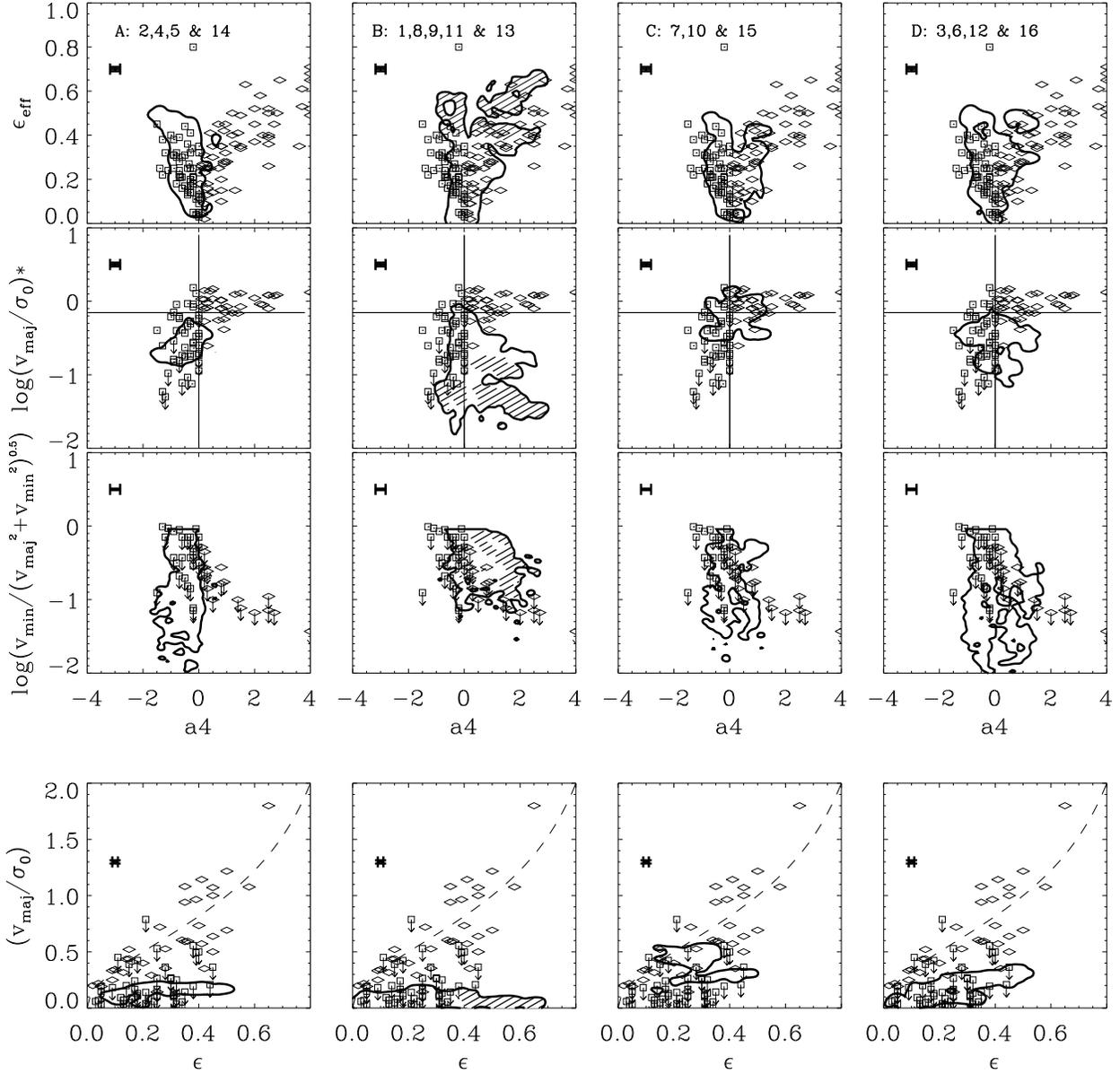}
\figcaption
{Statistical properties of equal-mass merger remnants divided into
  four groups (A,B,C,and D) according to their average properties (see
  text). The different initial geometries are given in the first row. The
  contours indicate the 90\% probability to find a projected remnant
  in the enclosed area. The shaded area in the second column
  corresponds to the location of projections with ellipticities larger
  than 0.4.\label{fig16}}  
\end{figure}

The present study will serve  as the basis of a further detailed
investigation on the influence of an additional dissipative
component. \citet{kb1996} proposed a revised 
Hubble sequence with disky ellipticals representing the missing link
between the Im-Spiral-S0 sequence and boxy ellipticals. They noted
that gas infall into the equatorial plane with subsequent star
formation could lead to  a second disk-like subcomponent. Ellipticals
with disks could appear disky when seen edge-on and
boxy otherwise \citep{sb1995}.  Our simulations indeed indicate that a
disk-like substructure is responsible for producing disky isophotes
(see Fig. \ref{fig6}). However, in the present case, the disk is the remnant of
the more massive spiral which was not completely destroyed during the
minor merger (see also \citet{b1998}). \citet{nb2001} investigated
line-of-sight velocity distributions of dissipationless 
merger remnants and found a velocity profile asymmetry that is opposite to
the observed one. They concluded that this disagreement can be solved
if ellipticals would contain a second disk-like substructure that most
likely formed through gas accretion. The situation is however not
completely clear, as another study by \citet{BB2000} found a good
agreement of the observed asymmetries for some cases.
\citet{nb2001} have shown that extended gas disks can form as a result
of a gas rich unequal-mass merger. \citet{bar2002} recently presented
a first detailed set of equal-and unequal-mass merger simulations of
gas-rich galaxies resulting in the formation of extended gas
disks. His simulations demonstrate a very complex dynamical evolution.
It has to be investigated how the presence of gas changes the detailed
properties of the remnants in detail. In addition, star formation
during the merger will influence the stellar kinematics (see
e.g. \citet{spr2000}) and the stellar populations by adding young and
probably more metal rich stars. As metal enrichment provides a futher
strong observational constraint on the formation history of early type
galaxies (see e.g. \citet{tmb2002}) more simulations, including
dissipation, star formation and chemical evolution will now be
required in order to understand in detail the role of gas for the
formation of elliptical galaxies by mergers.  

\acknowledgments

We thank Ralf Bender and Hans-Walter Rix for helpful comments on the
script and the referee, Joshua Barnes, for valuable suggestions that
significantly improved the paper. All the simulations presented in
this work have been performed on the GRAPE cluster at the
Max-Planck-Insitute for Astronomy in Heidelberg (MPIA). Thorsten Naab is
grateful to the MPIA for its support.

\clearpage

\begin{deluxetable}{ccccccc}
\tabletypesize{\scriptsize}
\tablecaption{Orbital parameters \label{tbl-1}}
\tablewidth{0pt}
\tablehead{
\colhead{Geometry\tablenotemark{a}} & \colhead{$i_1$}   & \colhead{$\omega_1$}   &
\colhead{$i_2$} &
\colhead{$\omega_2$}  & \colhead{$r_p$} & \colhead{$r_{\mathrm{sep}}$}}
\startdata
1/17 & 0 & 0       & 180& 0 & 2 & 30   \\
2/18 & 0 & 0       & 71& 30 & 2 & 30   \\
3/19 & 0 & 0       & 71& -30 & 2 & 30  \\
4/20 & 0 & 0       & 71& 90 & 2 & 30   \\
5/21 & -109 & -60  & 180 & 0 & 2 & 30  \\
6/22 & -109 & -60  & 71 & 30 & 2 & 30  \\
7/23 & -109 & -60  & 71 & -30 & 2 & 30 \\
8/24 & -109 & -60  & 71 & 90 & 2 & 30  \\
9/25 & -109 & 0    & 180 & 0 & 2 & 30  \\
10/26 & -109& 0    & 71 & 30 & 2 & 30  \\
11/27 & -109 & 0   & 71 & -30 & 2 & 30 \\
12/28 & -109 & 0   & 71 & 90 & 2 & 30  \\
13/29 & -109 & 60  & 180 & 0 & 2 & 30  \\
14/30 & -109 & 60  & 71 & 30 & 2 & 30  \\
15/31 & -109 & 60  & 71 & -30 & 2 & 30 \\
16/32 & -109 & 60  & 71 & 90 & 2 & 30  \\
 \enddata

\tablenotetext{a}{For unequal-mass mergers the first number indicates
  the orientation of the more massive galaxy as  $i_1$ and $\omega_1$,  
the second number indicates the orientation of the more massive
  galaxy as  $i_2$ and $\omega_2$. }
\end{deluxetable}

\end{document}